\documentclass[prl,twocolumn,aps,showpacs]{revtex4-1}

\usepackage{epsfig}
\usepackage{graphicx}
\usepackage{amsmath}
\usepackage{amsfonts}

\begin{document}

\title{Band-edge BCS-BEC crossover in a two-band superconductor: physical properties and detection parameters}

\author{Andrea Guidini$^{1}$ and Andrea Perali$^{2}$}

\affiliation{$^{1}$School of Science and Technology, Physics Division, University of Camerino, 62032 Camerino, Italy  \\ $^{2}$School of Pharmacy, Physics Unit, University of Camerino, 62032 Camerino, Italy\\}

\begin{abstract}
Superconductivity in iron-based, magnesium diborides, and other novel 
superconducting materials has a strong multi-band and multi-gap character. Recent experiments support the 
possibillity for a BCS-BEC crossover induced by strong-coupling and
proximity of the chemical potential to the band edge of one of the bands. 
Here we study the simplest theoretical model which accounts for the BCS-BEC 
crossover in a two-band superconductor, considering tunable 
interactions and tunable energy separations between the bands. 
Mean-field results for condensate fraction, correlation length, and superconducting gap are reported in typical crossover diagrams 
to locate the boundaries of the BCS, crossover, and BEC regimes. When the superconducting gap is of 
the order of the local chemical potential, superconductivity is in the 
crossover regime of the BCS-BEC crossover and the Fermi surface of the small 
band is smeared by the gap opening. 
In this situation, small and large Cooper pairs coexist in the total 
condensate, which is the optimal condition for high-T$_{\rm c}$ 
superconductivity. The ratio between the gap and the Fermi energy in a given band
results to be the best detection parameter for experiments to locate the system in the BCS-BEC crossover. Using available experimental data, our analysis shows that iron-based superconductors have the partial condensate of the small Fermi surface in the crossover regime of the BCS-BEC crossover,
supporting the recent ARPES findings.
\end{abstract}


\maketitle

\section{1. Introduction}
Multi-band and multi-gap superconductivity is emerging as a complex quantum coherent phenomenon with physical properties which are different or cannot be found in single band conventional superconductors. The increased number of degrees of freedom of the multi-component superconducting wave-function allows for novel effects. Phase solitons \cite{Lin2012} and massive or massless Leggett modes \cite{Leggett1966,Marciani2013} are possible benchmarks for multi-gap superconductivity, being associated with phase differences between the condensates in different electronic bands. Exotic vortex states \cite{Gillis2014}, non trivial interactions between the vortices \cite{Chaves2011}, and hidden criticality \cite{Komendova2012} are also peculiar phenomena associated with the multi-gap and multi-condensate nature of the superconducting state. In these systems, the total superconducting condensate results from the coherent mixture of partial condensates forming in each band, and the partial condensates can have very different properties, leading to interesting interference effects. Very recently, evidences of the BCS-BEC crossover and strong coupling superconductivity have been reported in the small Fermi surface pockets of multi-band supeconductors. ARPES measurements in iron-calchogenide superconductors have shown a superconducting gap to Fermi energy ratio of order unity in the shallow upper bands. The crossover regime has been detected by the collapse of the small Fermi surface pocket and by the electronic band dispersion becoming an inverted parabola in the coherent state \cite{Kanigel}. This phenomenology observed in iron-based superconductors is the same as the one predicted and observed in ultracold fermions \cite{Gaebler,Perali2011}.

Moreover, in underdoped superconducting cuprates the bandwidth around the M points of the Brillouin zone is very narrow, of the same order of the gap and pseudogap. Evidences of broken Fermi
surface and multi-band effects are also recently reported in YBCO superconducting cuprates due to charge ordering \cite{Shekhter2013}. A sizable pseudogap, its continuos evolution across T$\rm _c$, together with a short correlation length are evidences for strong pairing effects \cite{Norman2014} and importance of the BCS-BEC crossover in the underdoped cuprate superconductors \cite{Perali2000,Perali2002}.

Another motivation to study the BCS-BEC crossover in two-band superconductors comes from the superconducting properties of the iron-based superconductor Ba$_{0.6}$K$_{0.4}$Fe$_{2}$As$_{2}$ (T$\rm _c$=37K). Two different s-wave gaps open in the different sheets of Fermi surface (FS): a large gap of $\Delta_2$=12 meV on the small FS and a small gap $\Delta_1$=6 meV in the large FS \cite{Ding2008}. The ratio $2\Delta_1/ \rm T_c$=3.7
 is very close to the BCS value of 3.5, indicating BCS-like weakly coupled pairs in the large FS, while $2\Delta_2/ \rm T_c$=7.5 is very large and typical of BEC-like strongly coupled pairs in the small FS. Hence, the total superconducting condensate in BaKFeAs is a coherent mixture of BCS-like and BEC-like partial condensates.

A two-gap superconductor with quite distinct gaps in the different bands is also MgB$_2$. In this material evidences for resonant and crossover phenomena due to proximity to a band edge have been reported \cite{Innocenti2010}.

Quantum confinement and shape resonance in stripe systems, proposed as a mechanism for T$\rm _c$ amplifications \cite{Perali96} with recent experimental confirmation in metallic stripes \cite{Shanenko2006}, are clearly in coexistence with BCS-BEC crossover, which can determine the best situation for high-T$_{\rm c}$ superconductivity. Indeed, multi-band BCS-BEC crossover can determine the optimal condition to allow the screening of superconduciting fluctuations. This screening effect is expected to be active in a two-band superconductor, as shown by means of a Ginzburg-Landau approach in Ref.\cite{Perali2000}.

Quite generally, different families of iron-based superconductors show new small Fermi 
surfaces at optimum doping where T$_c$ is the highest,  appearing when
the chemical potential is near a band edge, close to the bottom (if electron
like) or top (if hole like) of the energy bands \cite{Borisenko2012}. 
In this situation, 
experiments show no evidences for nesting topology and the mechanism for high-T$_{\rm c}$ can be 
associated with the shape resonance scenario \cite{BianconiNP2013}.
In Fig. 1(a) and (b) of Ref.\cite{BianconiNP2013} the Fermi surfaces topology
for different superconducting iron-based materials have been schematized,
showing that in all cases large Fermi surfaces coexist with small Fermi 
surface pockets, supporting the (at least) two-band model for 
superconductivity as the minimal model to capture the band-edge physics
and corresponding novel multi-band BCS-BEC crossover phenomena. 

In 2D electron gases numerical evidences based on Bogoliubov-de Gennes equations for the BCS-BEC crossover in surface superconductivity have been found. In such systems the crossover phenomenon can be explored by controlling the gate voltage of the surface superconductivity. A gap to Fermi energy ratio larger than one signals the realization of the BEC regime in one of the bands of SrTiO$_3$, whereas the other bands are in the BCS regime having a gap to Fermi energy ratio smaller than one  \cite{Mizohata}.

Finally, in ultracold cigar-shaped Fermi gases the quantum confinement induces the formation of a series of single-particle subbands. As theoretically predicted in Ref.\cite{Shanenko2012}, in the superfluid state of these systems the total condensate is a coherent mixture of subband condensates, each of which undergoes a BCS-BEC crossover when the edge of the corresponding subband approaches the chemical potential.

The BCS-BEC crossover in two-band ultracold fermions in the superfluid state has been studied at mean-field and Ginzburg-Landau level in Ref.\cite{Sademelo}. Main results of Ref.\cite{Sademelo} have been the understanding of the role of the interband (Josephson-like) coupling in driving the transition between a 0-phase and a $\pi$-phase of the two component order parameter, the undamped collective excitations and the finite temperature Ginzburg-Landau description of the two-band BCS-BEC crossover.\\

In this work we will focus on the BCS-BEC crossover which can be induced in one of the partial condensates when the chemical potential is close to the band edge of a two-band system. In our model system, one partial condensate is in the weak-coupling regime with extedend Cooper pairs forming in the large Fermi surface, while the second partial condensate has tunable properties, and the pairing in the small Fermi surface can be varied from the weak- to the strong-coupling regime, allowing for the BCS-BEC crossover to be induced in the band associated with the small Fermi surface. We will explore at mean-field level and at zero temperature, in a three dimensional continuum, the phase space of the interaction parameters in order to detect the boundaries between the different BCS, crossover, and BEC regimes, locating in the boundary diagrams the available experimental data of multi-gap superconductors.\\
The physical description of the zero temperature
BCS-BEC crossover \cite{Eagles,Leggett} is based on two physical quantities: 
the condensate fraction \cite{Salasnich2005} and the average size of the Cooper pairs \cite{Pistolesi94}. 
The condensate fraction quantifies the fraction of fermions participating
in the condensation into the superconducting state: in the BCS regime
the condensate fraction is exponentially small and a very small density
of fermions forms the condensate wave-function; in the BEC regime 
the condensate fraction approaches unity, being the attraction so strong
that all the fermions form local molecular pairs with bosonic character.
In terms of the average pair size $\xi_{\rm pair}$ the physics is well known, being the
parameter $k_F\xi_{\rm pair}$ the first parameter used in the pioneering
work on the BCS-BEC crossover \cite{Pistolesi94}. The BCS regime is characterized by 
$k_F\xi_{\rm pair} \gg 1$, while in the BEC regime point-like pairs lead
to $k_F\xi_{\rm pair}<1$. This characterization of the BCS-BEC crossover
results to be very clear from a theoretical and phenomenological point of view, but unfortunately both the condensate fraction and the average size of the pairs are not easy to be measured. Hence, in this work we provide a way to link the BCS-BEC crossover given in terms of the condensate fraction $\alpha$ and $k_F\xi_{\rm pair}$ to the ratio between the superconducting pairing gap at $T=0$ to the Fermi energy of the non interacting system. This ratio is measurable in experiments by ARPES, tunnelling (STM) or specific heat measurements and can then provide a very useful parameter to detect the BCS-BEC crossover in complex multi-gap superconductors, as the case of iron-pnictides, MgB$_2$ or cuprates.\\

The paper is organized as follows. In Section 2 we will discuss the model system, the mean-field equations that describe it and the relevant physical quantities used to detect the boundaries of the BCS-BEC crossover. In Section 3 we will present our results for the one-band system as a reference, and then we will discuss the results for the two-band system in connection with available experimental data. Section 4 presents our concluding remarks.

\section{2. Model and methods}

\begin{figure}[tpb]
\centering
\includegraphics [angle=0,scale=0.7]{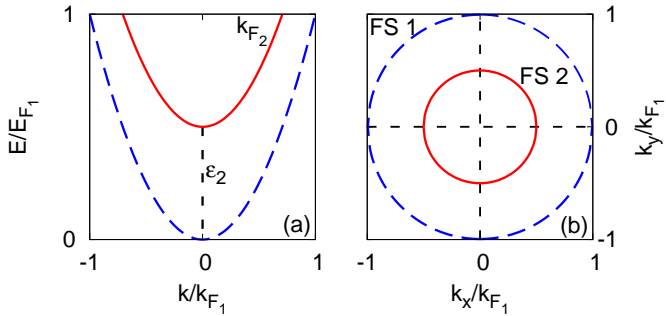}
\caption{Band dispersions (panel (a)) and $k_z$=0 projection of the Fermi surfaces (panel (b)). The energy and the wave-vectors are measured in units of $E_{F_1}$ and $k_{F_1}$, respectively. $\epsilon_2$ is the band offset of the upper band and it is measured in units of $E_{F_1}$.}
\label{band_fermi}
\end{figure}

A two-band system of interacting fermions in three dimensions and at zero temperature is
considered. The two electronic bands have a parabolic dispersion 
$\xi_i(\mathbf{k})$ given by:
\begin{equation}
\label{dispersion}
\xi_i(\mathbf{k})=\frac{\mathbf{k}^2}{2m}-\mu +\epsilon _i,
\end{equation}
being $\mathbf{k}$ the wave-vector, $m$ the effective mass which is taken equal in 
the two-bands, $\mu$ the chemical potential and $\epsilon_i$ the energy of the bottom of the bands. The index $i$=1,2 labels the bands: $i$=1 denotes the lower band and $i$=2 denotes the upper band. We set $\epsilon_1$=0, while $\epsilon_2 \geq$0 is tunable. The Fermi energies $E_{F_i}$ are defined in the non interacting case as $E_{F_i}=\mu - \epsilon_i$. We also set $\hbar$=1 throughout this article.\\
In Fig.\ref{band_fermi} the band dispersions and the Fermi surfaces are reported.
The effective pairing interaction between fermions is approximated by a
separable potential $V_{ij}(\mathbf{k},\mathbf{k}')$ with an energy cutoff $\omega_0$ and 
it is given by:
\begin{equation}
\label{potential}
V_{ij}(\mathbf{k},\mathbf{k}')=-V_{ij}^0\Theta\left(\omega _0 - |\xi_i(\mathbf{k})|\right)\Theta(\omega _0 - |\xi_j(\mathbf{k}')|),
\end{equation}
where $V_{ij}^0$ is the (positive) strength of the potential and $i$,$j$ label 
the bands. In the following we will set $V_{ij}^0=V_{ji}^0$. $V_{11}^0$ and $V_{22}^0$
are the strength of the intraband pairing interactions (Cooper pairs are created and 
destroyed in the same band). $V_{12}^0=V_{21}^0$
are the strength of the Josephson-like interband pairing interactions 
(Cooper pairs are created in one band and destroyed in the other band, and viceversa).
Here we consider the same energy cutoff $\omega_0$ of the interaction for intraband and interband pairing terms. Note that in this work we neglect interband pairing terms corresponding to Cooper pairs forming from fermions associated to different bands.
Moreover, when the chemical potential relative to the bottom of 
the bands becomes negative,
the replacement $\mu-\epsilon_i \rightarrow 0$ will be done in both the step functions of Eq.(\ref{potential}) 
to obtain the correct BEC limit given by the corresponding two-body problem 
in the vacuum. 
Indeed, when $\mu-\epsilon_i$ is negative the Fermi surface of the band ($i$) 
is destroyed and the interaction becomes contact-like, with a cutoff 
in $\mathbf{k}$ ($k_{c}=\sqrt{2m\omega_0}$) which ensures the convergence of the mean-field equations.\\

The superconducting ground state of our system is studied in this article at 
a mean-field level of approximation. We use mean-field equations at zero temperature 
based on the two-band extention of the mean-field BCS theory \cite{Suhl59}. 
The BCS equations for the two superconducting gaps are coupled with 
the equation for the total density of the system, being the renormalization of the 
chemical potential a key feature in the BCS-BEC crossover \cite{Perali2002,Pistolesi94}. \\

The equations for the gaps $\Delta _1(\mathbf{k})$ and $\Delta _2(\mathbf{k})$ read:
\begin{eqnarray}
\label{gap_eq_1}
\Delta _1(\mathbf{k})=&-&\frac{1}{\Omega} \sum_{\mathbf{k}'} \left[V_{11}(\mathbf{k},\mathbf{k}')\frac{\Delta _1 (\mathbf{k}')}{2\sqrt{\xi _1(\mathbf{k}')^2 +\Delta _1 (\mathbf{k}') ^2}}   \right. \nonumber \\ 
 &+& \left. V_{12}(\mathbf{k},\mathbf{k}')\frac{\Delta _2 (\mathbf{k}')}{2\sqrt{\xi _2(\mathbf{k}') ^2 +\Delta _2 (\mathbf{k}') ^2 }} \right],\\
\label{gap_eq_2}
\Delta _2(\mathbf{k})=&-& \frac{1}{\Omega} \sum_{\mathbf{k}'} \left[ V_{21}(\mathbf{k},\mathbf{k}')\frac{\Delta _1 (\mathbf{k}')}{2\sqrt{\xi _1(\mathbf{k}')^2+\Delta _1 (\mathbf{k}')^2}} \right. \nonumber \\ 
&+& \left. V_{22}(\mathbf{k},\mathbf{k}')\frac{\Delta _2 (\mathbf{k}')}{2\sqrt{\xi _2(\mathbf{k}') ^2 +\Delta _2 (\mathbf{k}')^2}}\right],
\end{eqnarray}
being $\Omega$ the volume occupied by the system, and with the gaps having the same cutoff of the separable interaction:
\begin{equation}
\label{gaps_k}
\Delta _i(\mathbf{k})=\Delta _i \Theta(\omega _0 - |\xi _i(\mathbf{k})|).
\end{equation}
Note that also the step function of Eq.(\ref{gaps_k}) undergoes the same replacement 
discussed above for the interaction potential in the case $\mu- \epsilon_i < 0$.\\

In this work, the total density of the two-band system is fixed and it is given 
by the sum of the densities in the two-bands, $n_{\rm tot}$=$n_1$+$n_2$.
The fermionic density $n_i$ in the band ($i$) at $T=0$ is defined as:
\begin{equation}
\label{density}
n_i=\frac{2}{\Omega}\sum_{\mathbf{k}}v_i(\mathbf{k})^2,
\end{equation}
where $v_i(\mathbf{k})$ is the BCS weight of the occupied states.
The BCS weights $v_i(\mathbf{k})$ and $u_i(\mathbf{k})$ are:
\begin{equation}
\label{vk2}
v_i(\mathbf{k})^2=\frac{1}{2}\left[ 1-\frac{\xi_i(\mathbf{k})}{\sqrt{\xi_i(\mathbf{k})^2+\Delta _i(\mathbf{k})^2}}  \right],
\end{equation}
\begin{equation}
\label{uk2}
u_i(\mathbf{k})^2=1-v_i(\mathbf{k})^2.
\end{equation}

The boundaries between the different BCS, crossover and BEC regimes
of the two-band BCS-BEC crossover are here determined by analyzing three 
fundamental physical quantities of the superconducting ground state wave function:
the condensate fraction, the correlation length of the Cooper pairs, and the 
superconducting gap, using physical insights from the BCS-BEC crossover in ultracold 
fermionic atoms close to a Fano-Feshbach resonance \cite{Pistolesi94,Salasnich2005}.
The condensate fraction $\alpha_i$, which is the ratio between the number of fermions 
of the band ($i$) forming the Cooper pairs of the condensate and the total number 
of fermionic particles in the same band, strictly related to the off-diagonal long-range order, is defined as:
\begin{equation}
\label{condfrac_eq}
\alpha _i=\frac{\sum_{\mathbf{k}}(u_i(\mathbf{k})v_i(\mathbf{k}))^2}{\sum_{\mathbf{k}} v_i(\mathbf{k})^2},
\end{equation}
while the pair correlation lenght $\xi_{{\rm pair},i}$, 
which represents the average size of the Cooper pairs is given by:
\begin{equation}
\label{csipair_eq}
\xi _{{\rm pair}, \phantom{} i}=\left[ \frac{\sum_{\mathbf{k}} |\nabla _{\mathbf{k}} (u_i(\mathbf{k})v_i(\mathbf{k}))|^2}{\sum_{\mathbf{k}} (u_i(\mathbf{k})v_i(\mathbf{k}))^2}\right]^{\frac{1}{2}}.
\end{equation}
In Eq.(\ref{csipair_eq}) the step function of Eq.(\ref{gaps_k}) is replaced by a smooth function in order to obtain finite partial derivatives.\\

We solve numerically the three by three system of Eqs. (\ref{gap_eq_1}), 
(\ref{gap_eq_2}) and (\ref{density}). We performe numerical calculations using a Fortran95 program in which we use the Gauss-Legendre method to evaluate the integrals over the energy variable $\epsilon=k^2/(2m)$ and the symmetric rank-one method to solve systems of equations. Once the values of the gaps and of the 
chemical potential are obtained, we calculate numerically the condensate 
fraction and the pair correlation length given in Eqs.(\ref{condfrac_eq}) and (\ref{csipair_eq}) respectively.\\ 
For the one-band system we use as units $k_F$ for wave-vectors, 
where $k_F=(6 \pi ^2 n)^{1/3}$ and $n$ is the density of a free Fermi gas. 
We use then the Fermi energy scale $E_F=\frac{k_F^2}{2m}$ to normalize the energies. 
The dimensionless coupling $\lambda$ is defined as $\lambda=N(E_F)V^0$ where $N(E_F)$ 
is the density of states at the Fermi level (note that in the one-band case we dropped 
the indices used in Section 2 to label the bands).\\ 
For the two-band system, as we are mostly interested in the properties 
of the upper band (label 2) we use the units of band 2 and energies will be 
measured in units of $E_{F_2}=E_{F_1}(1-\frac{\epsilon_2}{E_{F_1}})$, where $E_{F_1}$ is the Fermi energy of the lower band. The dimensionless 
coupling in the upper band is 
$\lambda_{22}^{eff}=\lambda_{22}\sqrt{1-\frac{\epsilon_2}{E_{F_1}}}$, 
with $\lambda_{22}=N(E_{F_1})V_{22}^0$. The interband (Josephson-like) coupling parameter are defined as $\lambda_{12}=N(E_{F_1})V_{12}^0$ and $\lambda_{21}=N(E_{F_1})V_{21}^0$. We set $\lambda_{12}=\lambda_{21}$.\\
The total density of fermions in our system 
is then $n_{\rm tot}=n_1\left[1+(1-\frac{\epsilon_2}{E_{F_1}})^{3/2}\right]$.\\

\section{3. Results}

The first step in our analysis is to study the properties
of the superconducting ground state with only one band and the 
separable interaction considered in this work. It turns out
that a full characterization of the BCS-BEC crossover for this fermionic system
is lacking in the literature even in the one-band case.
As described in the previous Section, the line boundaries between the 
BCS, crossover and BEC regimes
are determined through the calculation of the condensate fraction given in Eq.(\ref{condfrac_eq}), of the correlation length of the Cooper pairs 
given in Eq.(\ref{csipair_eq}), 
and of the superconducting gap obtained by reducing to the standard one-band case the Eqs.(\ref{gap_eq_1}-\ref{density}).
The aim is to verify that the crossover boundary lines  obtained with condensate fraction, correlation
length, and superconducting gap are compatible each other.\\

\begin{figure}[tpb]
\centering
\includegraphics [angle=0,scale=0.65]{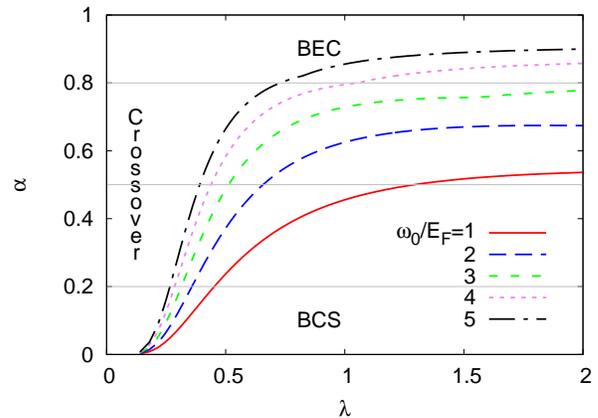}
\caption{Condensate fraction  $\alpha$ in the one-band case as a function of coupling $\lambda$ for different energy cutoffs of the interaction $\omega_0$ normalized to the Fermi energy. Thin solid lines (grey color on-line) correspond to $\alpha$=0.2, 0.5, 0.8 from bottom to top.}
\label{condfrac}
\end{figure}
In Fig.\ref{condfrac} the condensate fraction  $\alpha$ is reported 
as a function of the coupling
 $\lambda$ for different energy cutoffs of the pairing
interaction. This quantity will guide us to the exploration 
of the crossover boundary diagrams and to establish 
the boundaries for the gap and the pair correlation length. 
Thin horizontal lines (grey color on-line) represent our
choice of the boundaries between the different pairing regimes: 
for $\alpha<0.2$ the superconducting state is in the weak-coupling BCS
regime; for $0.2<\alpha<0.8$ the system is in the crossover regime; 
for $\alpha>0.8$ the system is in the strong-coupling BEC regime 
($\alpha=1$ corresponds to all fermions paired in a bosonic state).
The line $\alpha=0.5$ indicates the centre of the BCS-BEC crossover, 
correspoding to having $50\%$ of fermions in the condensed state.
Note that for $\lambda<0.25$ the condensate fraction, being directly 
proportional to the gap, is exponentially suppresed, and
the number of fermions entering in the condensate becomes extremely small. The proportionality between the gap and the condensate fraction in the BCS regime has been also obtained for ultracold fermions at mean-field level \cite{Salasnich2005}.
Note in Fig.\ref{condfrac} that when the 
pairing is increased the condensate fraction saturates to values that increase for increasing energy cutoff: as a consequence, while the crossover regime can be easily approached for different
cutoff values, the BEC regime requires strong paring and a large
energy cutoff to localize in space the Cooper pairs (large wave vectors need to be
coupled by the pairing interaction).\\

\begin{figure}[!h]
\centering
\includegraphics [angle=0,scale=0.65]{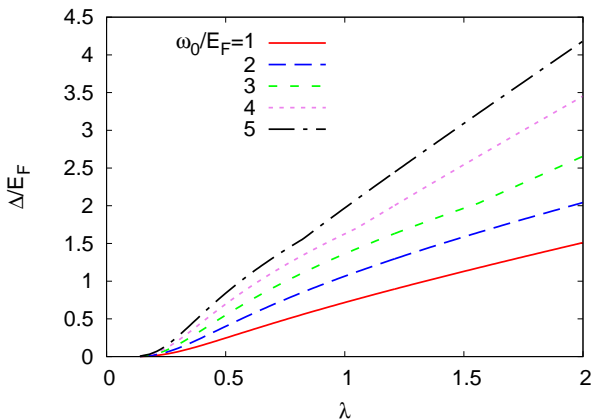}
\caption{Superconducting gap $\Delta$ in the one-band case as a function of the coupling $\lambda$ for different energy cutoffs of the interaction $\omega_0$ normalized to the Fermi energy.}
\label{delta}
\end{figure}
In Fig.\ref{delta} the superconducting gap in the one-band case is reported as a function 
of the coupling $\lambda$ for different values of the energy cutoff $\omega_0$. 
The gap is  exponentially suppressed for 
small values of the interaction $\lambda<$0.25 (the well known BCS 
weak-coupling limit $\Delta=2\omega_0e^{-1/\lambda}$ has been recovered by our numerical calculations) and it increases for 
larger value of the interaction when the BEC limit is approached. The gap gets larger and larger when 
the energy cutoff $\omega_0$ is increased. In the plot  for the coupling dependence of the gap for different energy cutoffs is not possible to find a single value of the gap which is in correspondence to the boundary values of the condensate fraction as shown later in this section.\\

\begin{figure}[!h]
\centering
\includegraphics [angle=0,scale=0.65]{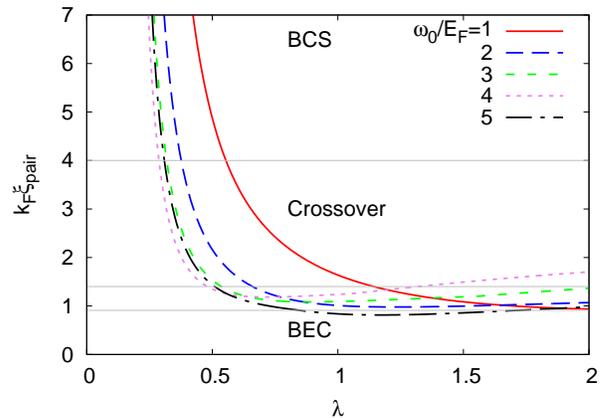}
\caption{Correlation length (average size of the Cooper pairs) $k_F\xi_{\rm pair}$ in the one-band case as a function of the coupling $\lambda$ for different energy cutoff of the interaction $\omega_0$ normalized to the Fermi energy. Thin solid lines (grey color on-line) correspond to $k_F\xi_{\rm pair}$=0.91, 1.4, 4.0 from bottom to top.}
\label{csipair}
\end{figure}
In Fig.\ref{csipair} the correlation length of the Cooper pairs is presented as a function 
of the coupling $\lambda$ for different values of the energy cutoff $\omega_0$. 
Thin grey lines represent the boundaries that we have found to match the crossover 
boundary diagram for $k_F\xi_{\rm pair}$ with that of $\alpha$ (see panel (c) 
of Fig.\ref{phase_oneband}). We obtain a satisfactory matching using $k_F\xi_{\rm pair}$=4.0 for the BCS boundary line, $k_F\xi_{\rm pair}$=1.4 for the center of the crossover line and $k_F\xi_{\rm pair}$=0.91 for the BEC line. The calculation of $k_F\xi_{\rm pair}$ requires to introduce a smearing procedure of the step function of Eq.(\ref{gaps_k}) in order to perform the numerical derivatives of Eq.(\ref{csipair_eq}). In Fig.4 the line corresponding to $\omega_0/E_F$=5 has been obtained with a different smearing parameter ($\omega_s/E_F$=4.0) with respect to the other four curves ($\omega_s/E_F$=0.1), and it has been presented to show how the BEC limit is approached.\\
In the weak-coupling limit ($\lambda \ll$1) the correlation length diverges, while 
it decreases for larger values of the pairing interaction. We have verified that in the weak-coupling limit ($\omega_0/E_F \lesssim 0.2$, $\lambda \lesssim 0.2$) our results for the pair correlation length are in good agreement to that of \cite{Pistolesi94}. Indeed we have found that our one-band system is in the BCS regime of pairing when the dimensionless parameter $k_F \xi_{\rm pair}>$4.0 which is close to the value $2\pi$ found in Ref.\cite{Pistolesi94} corresponding to strongly overlapping Cooper pairs. More preciseley, the boundary value for the BCS regime $k_F\xi_{\rm pair}=2\pi$
of Ref.\cite{Pistolesi94} is obtained in our model system for $\omega_0/E_F > 1$ and
for a value of the condensate fraction $\alpha$=0.13. Therefore, according to our choice of the boundary values, $k_F\xi_{\rm pair}=2\pi$ describes a regime of pairing well inside the BCS regime. Moreover, as expected from the 
behaviour of the gap and of the condensate fraction, also the pair correlation 
length approaches easily the crossover regime, while it approaches the BEC regime only for strong coupling and large energy cutoffs.

\subsection{Crossover boundary diagrams}
We present here the crossover boundary diagrams for the one-band 
system and for the two-band system. We consider the line boundary between the BCS and the crossover regime, the line corresponding to the center of the crossover, and the boundary line between the crossover and the BEC regime. In both one- and two-band systems, the constant values of the condensate fraction determine the crossover boundary lines. Along these boundaries we will extract the corresponding values of the pair correlation length and of the superconducting gap, which will be used as our detection parameters to locate the novel superconducting materials, as the iron-pnictides, in the BCS-BEC crossover phase boundary diagrams. In the case of the two-band system, our analysis of the BCS-BEC crossover will be focused on the upper band for chemical potentials close to the bottom of the upper band.\\

\subsubsection{3.1 One-band system}
\begin{figure}[h!]
\centering
\includegraphics [angle=0,scale=0.65]{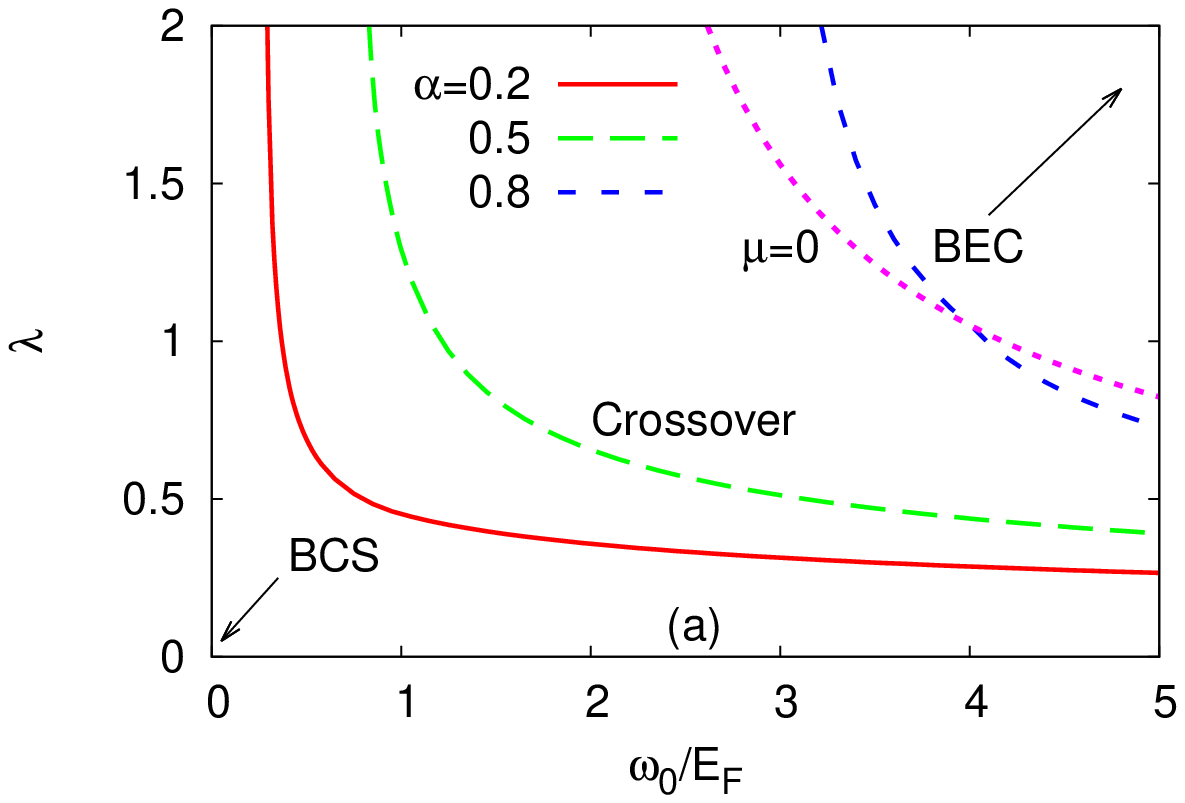}
\includegraphics [angle=0,scale=0.65]{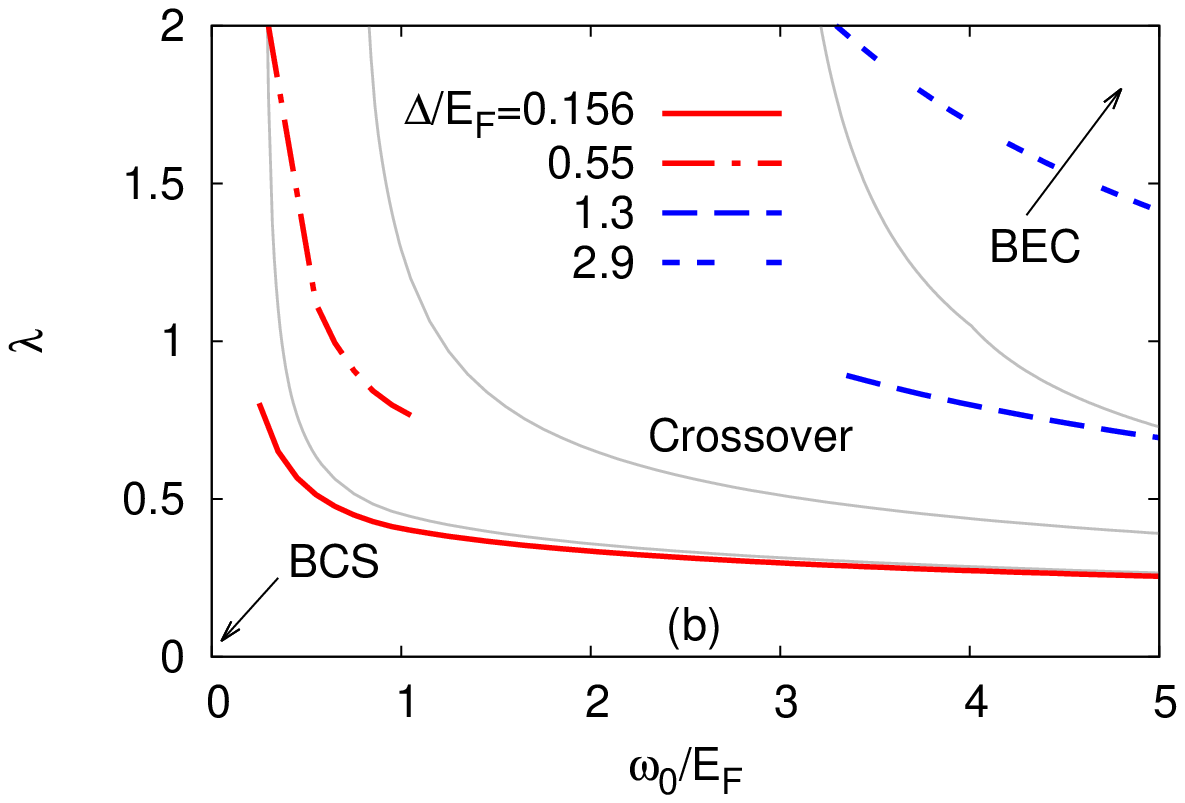}
\includegraphics [angle=0,scale=0.65]{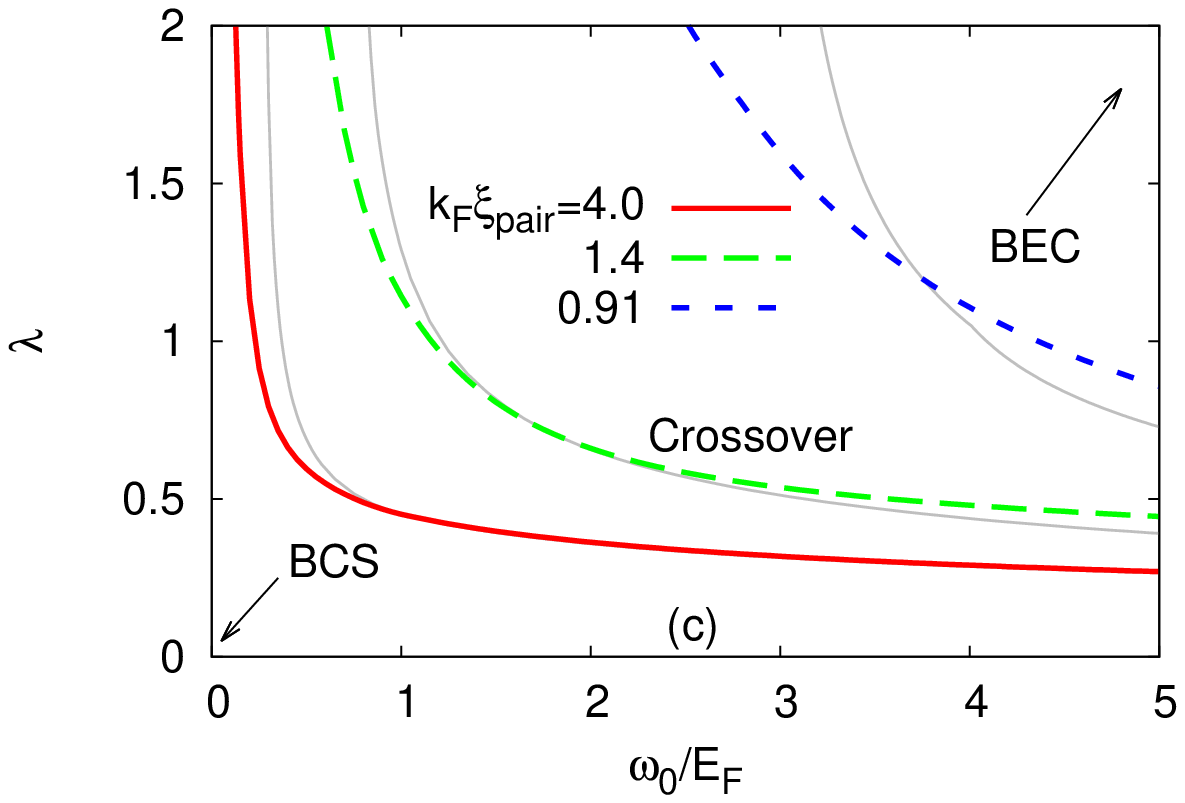}
\caption{Crossover boundary diagrams for the BCS-BEC crossover in the one-band case in the coupling vs energy cutoff plane. Different lines indicate the crossover boundaries between the BCS, crossover, and BEC regimes. The boundaries are obtained using the condensate fraction, the gap, and the correlation length. Panel (a): crossover boundary diagram for condensate fraction $\alpha$ and curve for $\mu=0$; panel (b): crossover boundary diagram for condensate fraction (thin solid lines, grey color on-line) and gap $\Delta$; panel (c): crossover boundary diagram for condensate fraction (same as in panel (b)) and pair correlation length $\xi_{\rm pair}$.}
\label{phase_oneband}
\end{figure}

Panels (a), (b) and (c) of Fig.\ref{phase_oneband} present the crossover 
boundary diagrams for the condensate fraction (panel (a)), gap (panel (b)) and 
pair correlation length (panel (c)) for the one-band system.\\
In panel (a) of Fig.\ref{phase_oneband} the crossover boundary diagram for 
the condensate fraction is presented together with the line that marks the change 
of sign of the chemical potential. In order to approach the BEC 
regime for the condensate fraction, it is necessary that the 
chemical potential becomes negative. This can be seen by comparing the boundary crossover 
lines for $\mu$=0 and $\alpha$=0.8. This is already an important result. From the physics of the BCS-BEC crossover in ultracold fermions, we know that $\mu$=0 corresponds to entering the BEC regime \cite{Perali2011}. Therefore we confirm that when more than 80\% of fermions are paired and condensed, the system is starting to be in the BEC regime.\\
Concerning the chemical potential we have found that in the BEC limit ($\lambda \gg 1 $ and $\omega_0/E_F \gg 1 $) the chemical potential $\mu$ approaches half of the binding energy of two particles which interact in the vacuum through the potential defined in Eq.(\ref{potential}). This result confirms the correct treatment of the BEC limit  that we have discussed in the previous section.
The equation that defines the (negative) binding energy $\epsilon_0$ of the two-body problem in the vacuum for a three dimensional system and the attractive interaction of Eq.(\ref{potential}) is:
\begin{equation}
\label{bound_state}
\frac{1}{\lambda}=\int_0^{\omega_0} d \epsilon \frac{\sqrt{\epsilon}}{2\epsilon + |\epsilon_0|}, 
\end{equation}
that has the implicit solution for $\epsilon_0$:
\begin{equation}
\label{bound_state_solution}
\frac{1}{\lambda}=\sqrt{\omega_0} - \sqrt{\frac{|\epsilon_0|}{2}} \arctan\sqrt{\frac{2\omega_0}{|\epsilon_0|}},
\end{equation}
where energy variables of Eqs.(\ref{bound_state}) and (\ref{bound_state_solution}) have been normalized to an energy scale $E_F$. In our many body system, we have verified that $\mu\rightarrow - |\epsilon_0|/2$ for sufficiently large values of the coupling $\lambda$ and of the energy cutoff $\omega_0$. For instance this BEC limit is obtained with our set of equations for $\omega_0/E_F \sim 6 $ and $\lambda \sim 3$. We have also found that there exists a critical value of the coupling $\lambda_c=1/\sqrt{\omega_0}$ (in units of a generic energy scale $E_F$) below which no two-body bound state is allowed: the existence of a coupling threshold for two-body bound state is a generic feature in 3D systems.\\\\

In panel (b) of Fig.5 the guiding crossover boundary diagram for the condensate fraction (thin grey lines) is 
presented together with the gap crossover boundary diagram. We choose the boundaries for the gap in order to match 
the condensate fraction crossover boundary lines at $\lambda$=2 and at $\omega_0/E_F$=5. 
We then choose $\Delta/E_F$=0.156 and $\Delta/E_F$=1.3 as lower bounds at $\omega_0/E_F$=5 for 
the BCS and the BEC lines respectively, and $\Delta/E_F$=0.55 and 2.9 as upper bounds for 
the gap at $\lambda$=2. Note that $\Delta/E_F$=0.156 is the value of the gap corresponding to $k_F \xi_{\rm pair}=$4.8.

In panel (c) of Fig.\ref{phase_oneband}, the guiding crossover boundary 
diagram for the condensare fraction is presented together with the correlation length crossover boundary diagram. 
The boundaries for the BCS, center of the crossover and BEC regimes are $k_F \xi_{\rm pair}$=4.0, 1.4 and 0.91 respectively. One can see that the BCS and 
center of the crossover boundary lines for the pair correlation length and for the condensate fraction are in good agreement for all the values
 of energy cutoff. The BEC line has the same qualitative behaviour of that for $\alpha$=0.8 and it is in very good agreement with the $\mu$=0 curve of panel (a). \\
To obtain the boundary lines for the correlation length 
we use a different smearing parameter per each boundary line: we choose $\omega_s/E_F$=0.1, 0.8 and 4 for the BCS, center of the crossover and BEC lines respectively.\\

To conclude the analysis of the BCS-BEC crossover in the one-band system we show in Fig.6 the crossover boundary diagram that presents the values of the gap to which correspond condensate fractions $\alpha$=0.2, 0.5 and 0.8, and $\mu=0$. The remarkable result that emerges from Fig.6 is that values of the gap between $\Delta/E_F$=0.55 and 1.3 permit to locate the system in the crossover regime of the BCS-BEC crossover for all the values of the energy cutoff $\omega_0/E_F$=(0,5) and of the coupling $\lambda$=(0,2) considered in this work. In Fig.6 we report for comparison the available experimental values of the ratio $\Delta/E_F$ reported in \cite{Ketterle} for ultracold fermionic atoms. The $\Delta/E_F$ data points are reported in Fig.6 with one arrow to indicate that the results of the contact potential can be obtained by using a separable attractive interaction only in the limit of large $\omega_0/E_F$ and small coupling $\lambda$.\\
\begin{figure}[!h]
\centering
\includegraphics [angle=0,scale=0.65]{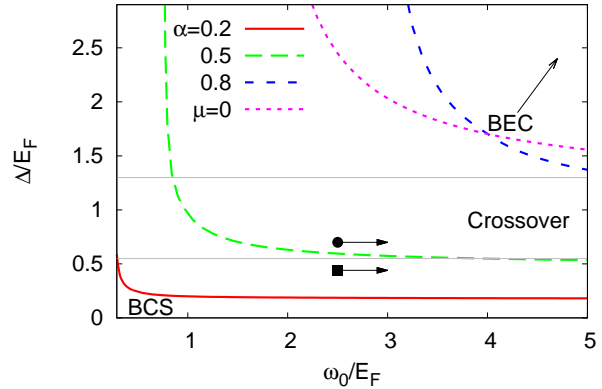}
\caption{Crossover boundary diagram in the one-band case for the gap $\Delta/E_F$ as a function of the cutoff frequency $\omega_0/E_F$ to which corresponds the condensate fractions $\alpha$=0.2, 0.5 and 0.8, and $\mu=0$. Thin horizontal lines (grey color on-line) represent the values $\Delta/E_F$=0.55 and 1.3 between which we have found that the system is in the crossover regime of pairing for every value of the coupling $\lambda$ and of the cutoff energy $\omega_0$ considered in this work. The points located at $\Delta/E_F$=0.44 and 0.70 correspond to two values of the pairing gap for ultracold fermions in the crossover regime reported in \cite{Ketterle}.}
\label{phase_delta}
\end{figure}
Using the crossover boundary diagram of Fig.\ref{phase_delta} one can identify immediately the values of the gap that permit to locate the superconducting system in the BCS, crossover or BEC regimes. Independently from the details of the pairing interaction, our results show that when $0.55<\Delta/E_F<1.3$ the one-band system is in the center of the BCS-BEC crossover and $\Delta/E_F$ can be considered as a robust detection parameter to characterize the regime of pairing.

\subsubsection{3.2 Two-band system}
\begin{figure}[!h]
\centering
\includegraphics [angle=0,scale=0.60]{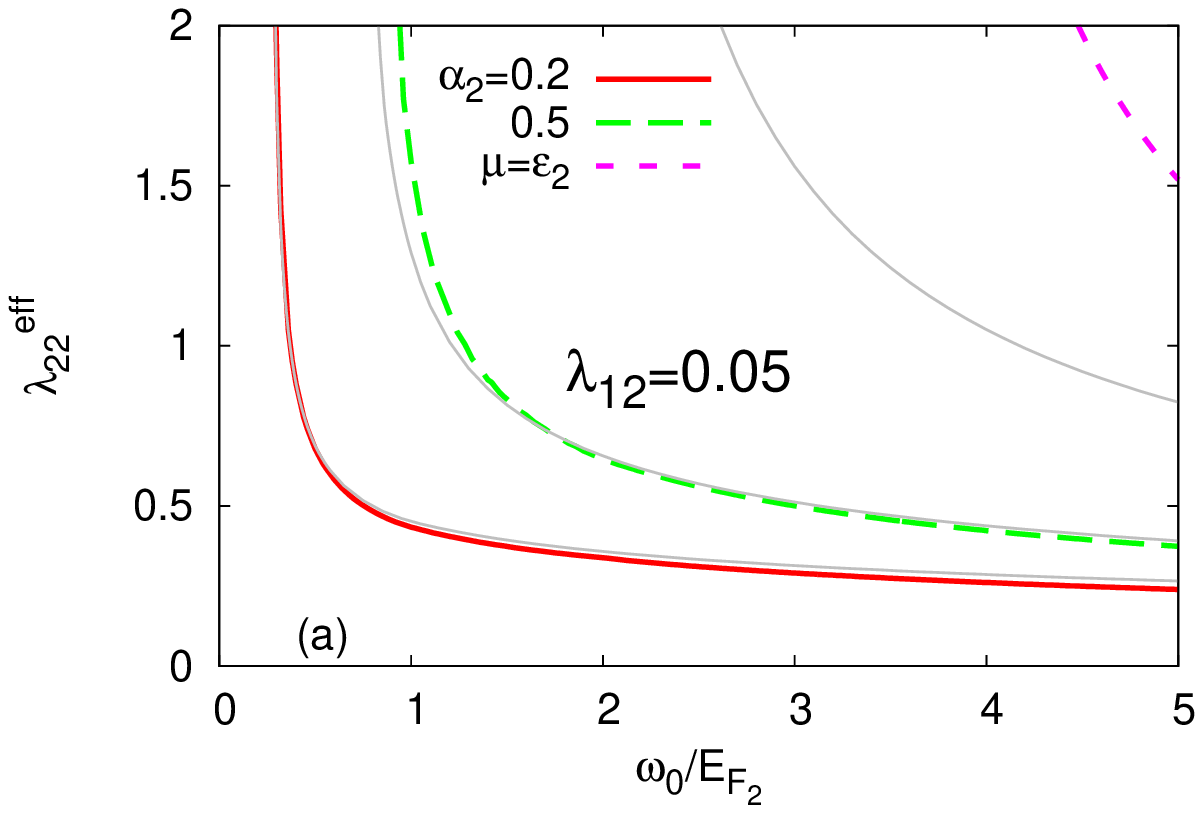}
\includegraphics [angle=0,scale=0.60]{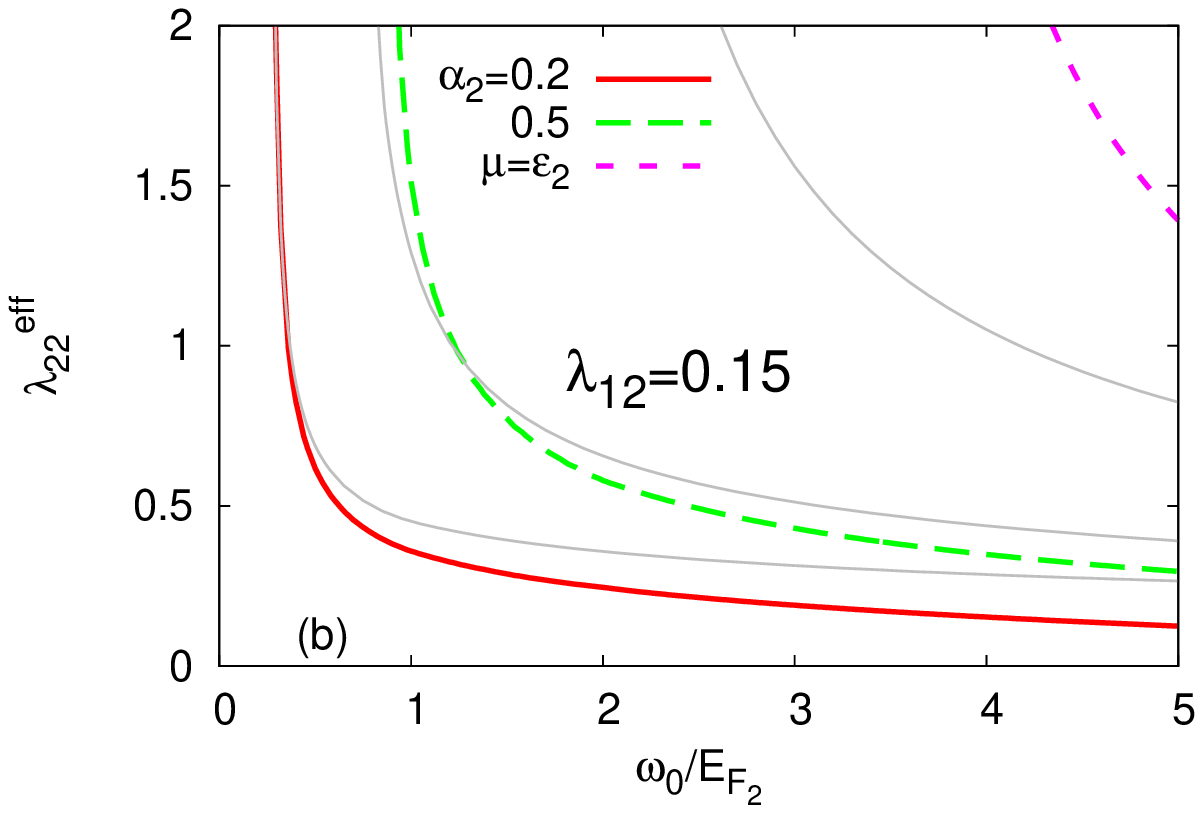}
\includegraphics [angle=0,scale=0.60]{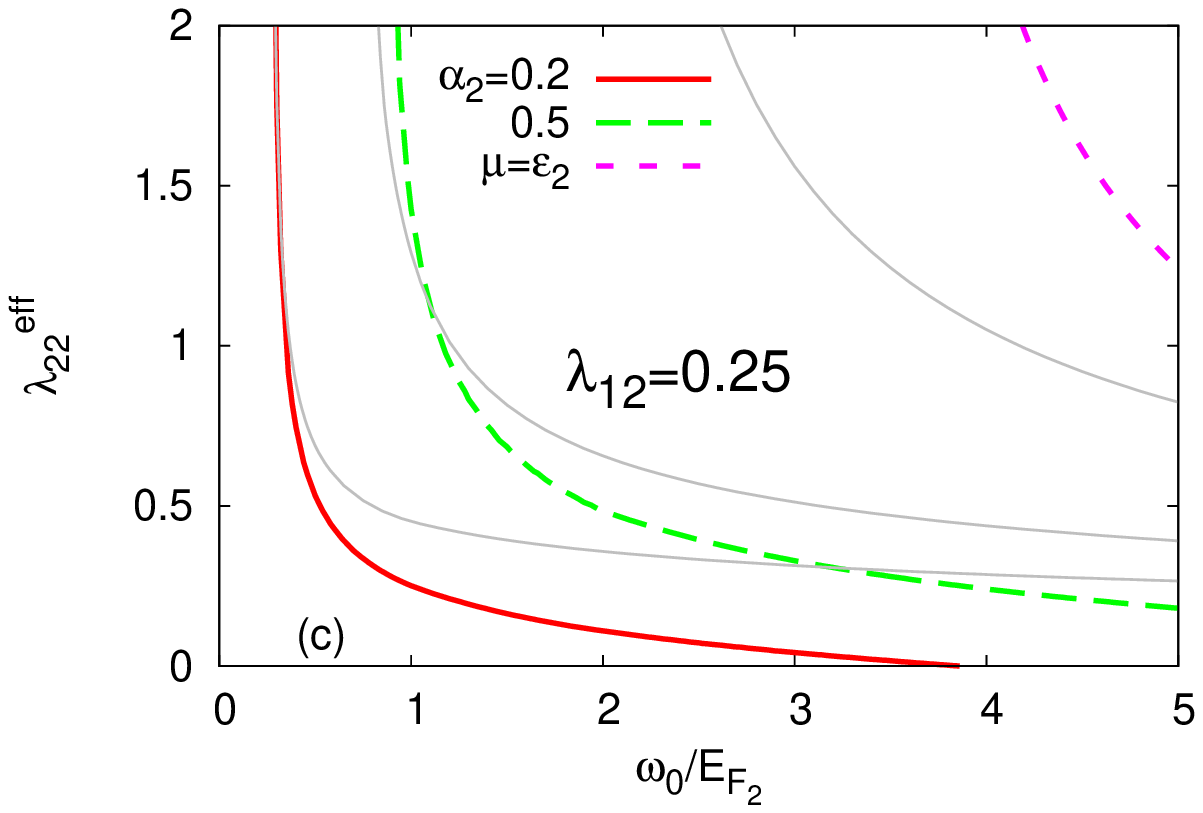}
\caption{Crossover boundary diagrams for the BCS-BEC crossover in the two-band case in the coupling vs energy cutoff plane for $\epsilon_2/E_{F_1}=0.5$, $\lambda_{11}=0.15$. The boundary lines for $\alpha_2$=0.2, 0.5 and $\mu=\epsilon_2$ for the two-band system are presented togheter with the lines for $\alpha$=0.2, 0.5 and $\mu=0$ for the one-band system (thin grey lines) as references. Different interband (Josephson-like) couplings are reported: $\lambda_{12}=0.05$ panel (a); $\lambda_{12}=0.15$ panel (b); $\lambda_{12}=0.25$ panel (c).}
\label{phase_twobands}
\end{figure}

Fig.\ref{phase_twobands} presents the crossover boundary diagrams for the partial condensate fraction in the upper band $\alpha_2$ and for the change of sign of the chemical potential with respect to the bottom energy of the upper band $\mu-\epsilon_2$=0. We consider the coupling $\lambda_{22}^{eff}$ vs energy 
cutoff $\omega_0/E_{F_2}$ plane for parameters: $\lambda_{11}$=0.15, $\epsilon_2$=0.5, 
for values of $\lambda_{12}=\lambda_{21}$=0.05, 0.15, 0.25 
(panels (a), (b) and (c) respectively). With the present choice of parameters the Cooper pairing in band 1 is in the weak coupling regime ($0.01<\Delta_1/E_{F_1}<0.5$), the interband pairing is progressively increased and the chemical potential is close to the upper band bottom, being $\mu-\epsilon_2$ on the same energy scale of the gap in band 2 and of the energy cutoff of the interaction.
In Fig.\ref{phase_twobands} the guiding lines for $\alpha$=0.2, 0.5 and $\mu$=0 (thin grey lines) of the one-band system (from panel (a) of Fig.5) are presented together with the crossover  boundary 
diagrams for the two-band system for $\alpha_2$=0.2, 0.5 and $\mu-\epsilon_2$=0. In panel (a) of Fig.7 corresponding to $\lambda_{12}=\lambda_{21}=0.05$, the BCS boundary line is almost not affected by the presence of the lower band whereas the center of the crossover boundary line is slightly retarded for $\omega_0/E_{F_2}\lesssim $1.5. In panel (b) and (c) of Fig.7 the BCS boundary line is not affected with respect to that of the one-band system for $\omega_0/E_{F_2} \lesssim $0.5 and progressively anticipated with increasing $\lambda_{12}$ for $\omega_0/E_{F_2} \gtrsim $0.5. The center of the crossover boundary line is slightly retarded and not affected by variation of $\lambda_{12}$ for $\omega_0/E_{F_2} \lesssim $1.1 and progressively anticipated for increasing $\lambda_{12}$ for $\omega_0/E_{F_2} \gtrsim 1.1$. For all the chosen values of $\lambda_{12}$ the value $\alpha_2$=0.8 to enter the BEC regime is not reached in the two-band system. Moreover in the two-band system the curve for 
the change of sign of the chemical potential with respect to the bottom of the upper 
band ($\mu-\epsilon_2$) is significantly retarded with respect to that of the one-band system. This is because the presence of the lower 
band (having a gap smaller than the gap in the upper band) locks the chemical potential 
and does not permit it to decrease as fast as in the one-band system. As a consequence, the boundary line $\mu=\epsilon_2$ is reached only for very large values of $\omega_0/E_{F_2}$ and $\lambda_{22}^{eff}$. The line $\mu-\epsilon_2$=0 is progressively anticipated for increasing values of $\lambda_{12}$.\\
Note that the line for $\alpha_2$=0.2 in panel (c) vanishes at $\omega_0/E_{F_2} \sim $4 meaning that the system is already out of the BCS regime also with zero intraband coupling in band 2 because of the sizeable interband coupling $\lambda_{12}$.\\

\begin{figure}[!h]
\centering
\includegraphics [angle=0,scale=0.70]{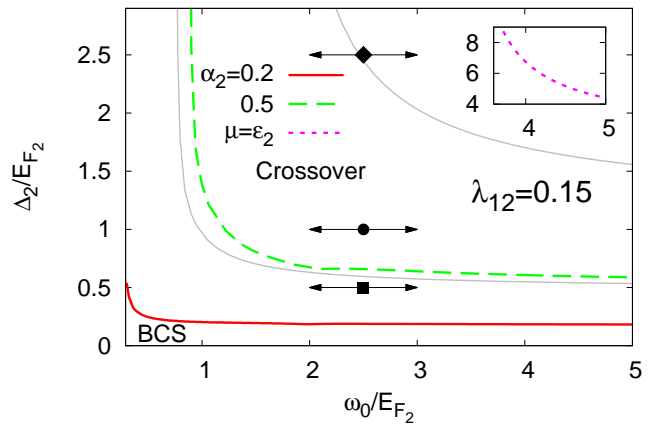}
\caption{Crossover boundary diagram for the gap $\Delta_2/E_{F_2}$ as a function of the cutoff frequency $\omega_0/E_{F_2}$ to which correspond condensate fractions $\alpha_2$=0.2, 0.5 and chemical potential $\mu=\epsilon_2$ in the two-band system with $\epsilon_2$=0.5, $\lambda_{11}=\lambda_{12}=\lambda_{21}$=0.15. The lines for the one-band system (thin grey lines) for the gap $\Delta/E_F$ to which correspond condensate fractions $\alpha$=0.2, 0.5 and chemical potential $\mu=0$ are also reported for reference. Inset: value of the $\Delta_2/E_{F_2}$ when $\mu=\epsilon_2$. The points $\Delta_2/E_{F_2}$=0.5, 1.0 and 2.5 correspond respectively to the experimental values reported in \cite{Kanigel}, \cite{Okazaki} and \cite{Miao}.}
\label{phase_delta_duebande}
\end{figure}
In Fig.\ref{phase_delta_duebande} the boundary crossover diagrams reporting the values of the gap $\Delta_2/E_{F_2}$ that give $\alpha_2$=0.2, 0.5 and $\mu-\epsilon_2$=0 (inset) are presented for the two-band system using the same parameters of panel (b) of Fig.\ref{phase_twobands}, together with the lines for the one-band system presented in Fig.6 (we present in Fig.8 the lines for $\alpha$=0.2, 0.5 and  $\mu$=0 for the one-band system). This is the central result of our work for the two-band system.\\
One can see in Fig.\ref{phase_delta_duebande}
 that the BCS gap line of the two-band system corresponding to a partial condensate fraction in band 2 of $\alpha_2$=0.2 perfectly overlaps the line of the one-band system, whereas the center of the crossover line is slightly shifted up. The BEC line $\alpha_2$=0.8, as we said above, is not reachable in the two-band system, whereas the line for $\mu=\epsilon_2$ is out of the chosen range ($\Delta_2/E_{F_2} \sim 6$ presented in the inset). In Fig.8 we report for comparison the available experimental values of the ratio $\Delta_2/E_{F_2}$ for iron-based superconductors measured by ARPES. In the case of FeSe$_x$Te$_{1-x}$ the gap to $E_F$ ratio in the small upper band is $\Delta_2/E_{F_2}\approx$ 0.5 for $x=$0.35 \cite{Kanigel} while for $x=0.40$ is $\Delta_2/E_{F_2}\approx$1.0 \cite{Okazaki}. On the other hand very recent ARPES results give $\Delta_2/E_{F_2}\approx$2.5 for LiFe$_{1-x}$Co$_x$As ($x$=0) and a negative chemical potential with a finite gap for the same compound with 1$\%$ and 3$\%$ Co-doping \cite{Miao}. The $\Delta_2/E_{F_2}$ data points are reported in Fig.8 with two arrows to indicate that the characteristic energy scale of the pairing interaction is not known, but it is of the order or larger than the (local) Fermi energy of the upper band, having the pairing an electronic origin.\\
Therefore, the central result of our work is that: ($i$) the superconducting iron chalcogenides are in the middle of the crossover regime of the BCS-BEC crossover, while ($ii$) Co-doped iron-pnictides superconductors are close to the BEC regime, with the BEC character being more pronounced for increasing Co-doping.

Our conclusion is that the two-band system with moderate intraband coupling for $0.55<\Delta_2/E_{F_2}<4$ is in the crossover regime of the BCS-BEC crossover for any choice of pairing $\lambda_{22}^{eff}$ and cutoff frequency $\omega_0/E_{F_2}$ considered in this paper. As a consequence the crossover region of the two-band system is wider than that of the one-band system. On the other hand, only extreme values of $\Delta_2/E_{F_2}>$4 and very large values of $\omega_0/E_{F_2}$ (plausible for pairing glue having an electronic origin) will drive the system toward the BEC regime for the partial condensate in the upper band.\\

We now pass to discuss our results on the average size of the Cooper pairs in the two bands ($k_{F_1}\xi_{ {\rm pair},1}$ and $k_{F_2}\xi_{ {\rm pair},2}$). The ratio between the two gaps $\Delta_2/\Delta_1$ will be also reported being a quantity directly comparable with experiments. We consider here two cases in the crossover regime in Fig.8:\\
($i$) for $\Delta_2/E_{F_2}$=1 and $\omega_0/E_{F_2}$=2.5 we obtain $k_{F_1}\xi_{ {\rm pair},1}$=8.4 and $k_{F_2}\xi_{ {\rm pair},2}$=1.2 with a ratio $\Delta_2/\Delta_1$=4. In this situation we have extended overlapping Cooper pairs in band 1 and molecular-like bosonic pairs in band 2. The total condensate is a coherent mixture of Cooper pairs and (almost) point-like molecular pairs.\\
($ii$) for $\Delta_2/E_{F_2}$=0.5, same $\omega_0/E_{F_2}$=2.5, the results are less extreme. We obtain $k_{F_1}\xi_{ {\rm pair},1}$=10.6 and $k_{F_2}\xi_{ {\rm pair},2}$=2.2, with a ratio $\Delta_2/\Delta_1$=2.5. The Cooper pairs in band 2 start to overlap and the corresponding partial condensate is in the middle of the crossover regime between BCS and BEC (indeed in Fig.8 the experimental point $\Delta_2/E_{F_2}$=0.5 is very close to the line $\alpha$=0.5 indicating the center of the crossover regime). This second case is very close to the experimental findings for the superconductor Ba$_{0.6}$K$_{0.4}$Fe$_2$As$_2$ as reported in Ref.\cite{Ding2008} ($\Delta_1$=6 meV, $\Delta_2$=12 meV).

We finally conclude with an observation concerning the results that we obtain increasing the bottom band energy of the upper band. For $\epsilon_2/E_{F_1}$=0.7 (and same choice of $\lambda_{11}$=0.15, $\lambda_{12}=\lambda_{21}$=0.05, 0.15 and 0.25) the BCS boundary line and the center of the crossover line for the partial condensate in the upper band ($\alpha_2$=0.2 and 0.5 respectively) are almost not affected by the different choice of $\epsilon_2/E_{F_1}$, being also the corresponding values of the gap not much affected. On the other hand the boundary line $\mu-\epsilon_2$=0 is placed to larger values of the coupling $\lambda_{22}^{eff}$ and of the cutoff frequency $\omega_0/E_{F_2}$, and the corresponding gap $\Delta_2/E_{F_2} \sim $7 is also larger than in the case $\epsilon_2/E_{F_1}$=0.5. One possible reason for this (unexpected) result is that the increased separation $\epsilon_2$ between the bands leads to a larger density in band 1, determining a slower variation of the chemical potential $\mu$ for increasing coupling.

%%%%%%%%%%%%%%%%%%%%%%%%%%%%%%%%%%%%%%

%%%%%%%%%%%%%%%%%%%%%%%%%%%%%%%%%%%%%%

\section{4. Conclusions}
We have analyzed the ground state superconducting properties of one- and two-band systems of fermions interacting through a separable attractive potential with an energy cutoff.\\
For the one-band system we have analyzed  the behaviour of the superconducting gap, condensate fraction and pair correlation length as a function of coupling for different values of the cutoff. We have found that our model system gives expected results both in the weak- and in the strong-coupling limits. The pair correlation length and the superconducting gap recover the BCS limit for weak interactions, and the chemical potential approaches half of the binding energy of a bound state of two fermions in the BEC limit. By choosing three values for the condensate fraction we have defined the boundaries of the BCS, crossover and BEC regime of the BCS-BEC crossover, and we have determined the corresponding boundary values for the superconducting gap and for the pair correlation length. The superconducting gap to Fermi energy ratio is the most accessible physical quantity that can be measured in experiments. We have found that when the ratio between the gap and the Fermi energy is in the range $0.55<\Delta/E_F<1.3$ the system is in the crossover regime of the BCS-BEC crossover for every value of the coupling and of the energy cutoff. Therefore this connection between $\Delta/E_F$ and the crossover regime is not associated with the particular microscopic structure of the pairing interaction. The ratio $\Delta/E_F$ results to be a useful detection parameter to characterize the regime of pairing of a superconductor, locating it in the BCS-BEC crossover.\\
For the two-band system we have analyzed the crossover boundary diagrams for the partial condensate fraction in the upper band and for the change of sign of the chemical potential relative to the bottom of the upper band. Different values of the interband coupling have been considered, assuming a weak-coupling regime of pairing in the lower band. We found that for increasing values of the interband coupling the access to the crossover regime is progressively anticipated with respect to the one-band case. The BEC regime for the partial condensate fraction is not reached in our range of coupling and energy cutoff, because the presence of the lower band locks the chemical potential, retarding the  change of sign of the chemical potential relative to the bottom of the upper band, and preventing the two-band system from accessing the BEC regime, widening in this way the crossover regime of the BCS-BEC crossover.\\

In conclusion, our work gives a simple and quite general theoretical support to the  recent ARPES measurements in 
iron-based superconductors which provide evidences that the Cooper pairs in the small Fermi surface form a condensate which is at the border between the crossover and the BEC regime. Therefore, the BCS-BEC crossover seems to be clearly realized in this new class of high-T$\rm _c$ superconductors.\\

\begin{acknowledgments}
We are grateful to A. Bianconi, F. Marsiglio, P. Pieri and L. Salasnich for useful discussions. We acknowledge financial support from 
the University of Camerino under the project FAR ``Control and enhancement of superconductivity by engineering materials at the nanoscale''.
\end{acknowledgments}

%%%%%%%%%%%%%%%% REFERENCES %%%%%%%%%%%%%%%%%%%%%%%%%%%%%%

\end{document}